# Magnetic nanowires generated via the waterborne desalting transition pathway

**M. Yan[a], J. Fresnais[b], S. Sekar[c], J.-P. Chapel[c] and J.-F. Berret[a]\***

[a]Matière et Systèmes Complexes, UMR 7057 CNRS Université Denis Diderot Paris-VII, Bâtiment Condorcet, 10 rue Alice Domon et Léonie Duquet, 75205 Paris (France)
[b] Physicochimie des Electrolytes, Colloïdes et Sciences Analytiques (PECSA) UMR 7195 CNRS-UPMC-ESPCI, 4 place Jussieu, 75252 Paris Cedex 05
[c]Centre de Recherche Paul Pascal (CRPP), UPR CNRS 8641, Université Bordeaux 1, 33600 Pessac - France

**Abstract :** We report a simple and versatile waterborne synthesis of magnetic nanowires following the innovative concept of electrostatic ''desalting transition''. Highly persistent superparamagnetic nanowires are generated from the controlled assembly of oppositely charged nanoparticles and commercially available polyelectrolytes. The wires have diameters around 200 nm and lengths comprised between 1 μm and ½ mm, with either positive or negative charges on their surface. Beyond, we show that this soft-chemistry assembly approach is a general phenomenon independent of the feature of the macromolecular building blocks, opening significant perspectives for the design of multifunctional materials.

\*Corresponding author. Email: jean-francois.berret@univ-paris-diderot.fr

# I – Introduction

Magnetic nanowires have received considerable attention during the past years because of their potential applications in cell manipulation[1-5], microfluidics[6] and micromechanics[7-9]. These objects are anisotropic colloidal particles with submicronic diameters and lengths in the range 1 – 100 μm. For specific applications, elongated magnetic structures appear as promising alternatives to the spherical magnetic beads developed so far. At the application of a magnetic field, torques may be applied to the wires and their orientational behavior be monitored, providing additional degrees of freedom for investigating complex fluids. Wires can also be made hollow with an internal surface in contact with the bulk phase and be used for separation and release of actives[10].

As far as the synthesis of magnetic nanowires is concerned, two main strategies are put forward in the literature. The first approach consists in the electrodeposition of magnetic atoms such as nickel or iron into thin alumina-based porous templates with cylindrical holes[11,12]. At the dissolution of the template, ferromagnetic nanowires are produced and dispersed in water-based solvents. These wires are used in cell guidance, cell separation and microrheology experiments[2-4,6,12]. One major drawback with nickel nanowires is that these structures are carrying a permanent magnetic moment, and thus aggregate spontaneously due to magnetic dipolar interactions[4,5]. The second strategy is based on the co-assembly of polymer microbeads loaded with magnetic nanoparticles[7-9,13,14]. With this technique, permanent chains are formed by the application of a magnetic field for alignment and by the addition of binding agents for connecting the beads. Filaments and chains with micronic diameters and length $1 - 200$ μm are fabricated, either in bulk solutions[8,13] or tethered on substrates[7,14,15]. The two synthesis routes end up however with different magnetic structures : the nickel nanowires are stiff and ferromagnetic, whereas the bead assemblies remain paramagnetic and semiflexible. Attempts to build supracolloidal assemblies using sub-10 nm particles were also carried out, resulting in isotropic clusters[16-18] or in tortuous pearl-necklace filaments[19-21].

We report here a simple and versatile waterborne synthesis of magnetic nanowires following the innovative concept of electrostatic ''desalting transition'' put forward recently on a sub-10 nm nanoparticles/charged-neutral block copolymers system[22,23]. We generalize this electrostatic co-assembly pathway to strong and weak homopolyelectrolytes (homoPEs) and demonstrate that the transition between a disperse and an aggregated state of particles observed as a function of the ionic strength is a general phenomenon independent of the nature of the polyelectrolytes considered. Furthermore, under the presence of a magnetic field this soft chemistry approach enables the fabrication *on-demand* of highly persistent superparamagnetic nanowires with lengths comprised between 1 μm and ½ mm bearing either positive or negative charges on their surface (Fig. 1).

## II – Materials and Methods

Iron oxide nanoparticles were synthesized according to the Massart technique [24] by alkaline co-precipitation of iron(II) and iron(III) salts, oxidation of the magnetite ($Fe_3O_4$) into maghemite ($\gamma$-$Fe_2O_3$) nanoparticles and by size-sorting by subsequent phase separations [25]. At the end of the synthesis, the particles were positively charged and dispersed in water at a weight concentration c ~ 10 wt. % and pH 1.8. The magnetic dispersions were characterized by transmission electron microscopy (TEM) and diffraction, vibrating sample magnetometry (VSM), magnetic sedimentation and light scattering (DLS). For the dispersions investigated here, the size distribution was found to be log-normal with median diameter 8.3 nm and a polydispersity 0.26 [26]. The polydispersity index is defined as the ratio between the standard deviation and the average diameter. To improve their stability, the particles were coated with $M_W$ = 2000 g mol$^{-1}$ poly(acrylic acid) using the *precipitation-dispersion* technique [26]. The thickness of the PAA$_{2K}$ brush was estimated at 3 nm by dynamical light scattering, corresponding to 25 ± 3 chargeable carboxylic groups per nm$^2$. A complete characterization of the $\gamma$-Fe$_2$O$_3$ nanoparticles with and without coating is provided in Supporting Information (SI.1).

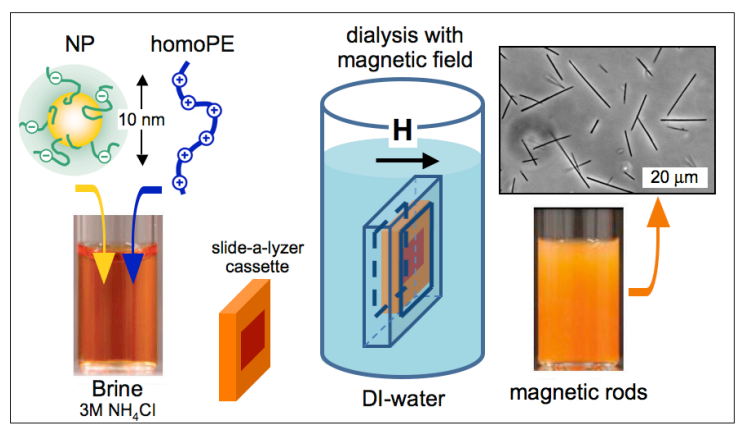

**Figure 1 :** *Schematic representation of the desalting transition protocol that controls the nanoparticle co-assembly and wire formation.*

In a previous work, electrostatic complexation schemes were reported using charged–neutral diblock copolymers synthesized by controlled radical polymerization (MADIX®) [27]. The polymer was poly(trimethylammonium ethylacrylate)-*b*-poly(acrylamide) with molecular weight 11000 g mol$^{-1}$ and 30000 g mol$^{-1}$ for each block, abbreviated as PTEA$_{11K}$-*b*-PAM$_{30K}$ [28]. Here, we generalized this strategy to widespread homoPEs (Sigma), including $M_W$ < 100000 g mol$^{-1}$ poly(diallyldimethylammonium chloride), 2000 g mol$^{-1}$ poly(ethyleneimine) and 15000 g mol$^{-1}$ poly(allylamine hydrochloride). The cationic homoPEs used here are typical building blocks encountered in the fabrication of multilayers polyelectrolytes via the layer-by-layer sequential adsorption technique [29,30]. In the following, they will be abbreviated as PDADMAC, PEI and PAH respectively (SI-2). In the mixed polymer/particle solutions, the relative amount of each component was monitored by the charge ratio Z. Z was defined as the ratio between the cationic charges borne by the polymers and the anionic charges carried by the particles at their surfaces. With this notation, Z = 1 describes the isoelectric solution, characterized by the same number densities of positive and negative chargeable ions. For the desalting transition pathway, ammonium chloride (Sigma) was used to adjust the ionic strenght up to 3 M.

# III – Results and discussion

The aggregate formation was monitored using different formulation pathways, including *direct mixing*, *dialysis* and *dilution*. Only *dialysis* was performed in the presence of an external applied magnetic field, in agreement with the scheme of Fig. 1.

## III.1 - Direct mixing

With the *direct mixing* protocol, the complexation of oppositely charged species was controlled by pouring rapidly the magnetic dispersion into the polymer solution at the desired charge ratio Z. The samples were allowed to rest for a week before being investigated by light scattering. Fig. 2a displays the hydrodynamic diameters ($D_H$) obtained for $PAA_{2K}-\gamma-Fe_2O_3$ complexed with $PTEA_{11K}-b-PAM_{30K}$ copolymers and with PDADMAC, PAH and PEI polycations. For the 4 systems, the $D_H$'s were found to pass through a sharp maximum around Z = 1. The measured diameters for $PTEA_{11K}-b-PAM_{30K}$ diblocks ($D_H = 100$ nm) were lower than those of the homoPEs ($D_H = 300$ nm), indicating a softening of the assembly process when non-interacting neutral blocks are present. The intense peaks around the charge stoichiometry bear strong similarities with that observed for polyplexes and lipoplexes in transfection strategies, suggesting analogous complexation mechanisms[31-33].

## III.2 – Desalting transition without magnetic field

Although appealing and simple, the *direct mixing* protocol was not appropriate to control the sizes and morphologies of hybrid nanostructures. Indeed, under strong driving forces like electrostatic interactions, the key factor in the complex formation is the competition between the reaction time of the components and the homogenization time of the solutions[34,35]. To overcome these limitations, alternative pathways inspired from molecular biology were developed[22,23]. The protocols applied here consist first in the screening of the electrostatic interactions by bringing the dispersions to high ionic strength ($I_S$), and second in the progressive reduction of $I_S$ by *dialysis* (Fig. 1) or by *dilution*. With this technique, the oppositely charged species were intimately mixed in solution but did not interact owing to the electrostatic screening. In the dilution process, deionized water was added stepwise to the polymer/particle salted dispersion and the scattered intensity and hydrodynamic diameter were determined by light scattering. For a dispersion containing both anionic particles and cationic $PTEA_{11K}-b-PAM_{30K}$ diblocks, $D_H$ *versus* $I_S$ exhibited an abrupt upturn (Fig. 2b). Found at 0.37 M, this upturn sets the limit between two states : above, the particles remain disperse and unaggregated; below, the particles are retained within dense spherical clusters thanks to the cationic polymer "glue"[22]. No transition could be evidenced for the particles or for the copolymers taken separately (SI-3).

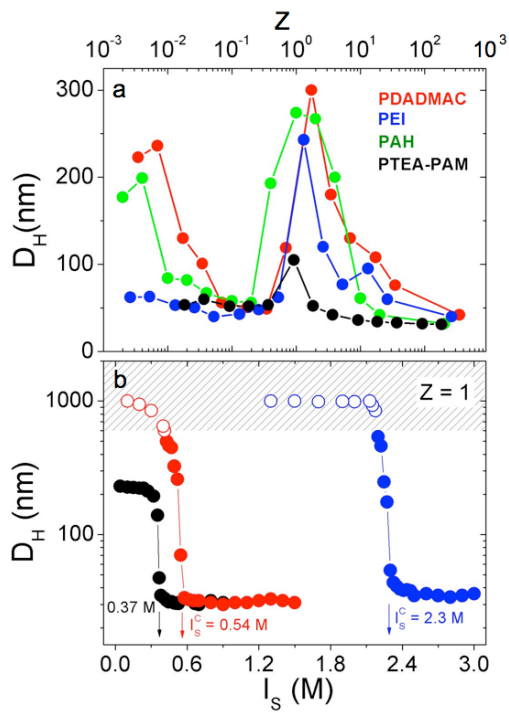

**Figure 2 : (a)** *Hydrodynamic diameter $D_H$ as a function of Z for $PAA_{2K}-\gamma-Fe_2O_3$ complexed with homo- and copolymers for charge ratios $10^{-3} - 100$ (black : $PTEA_{11K}-b-PAM_{30K}$; red : PDADMAC; blue : PEI and green : PAH). The polymer/particle dispersions were obtained by direct mixing at concentration 0.1 wt % and temperature 25 °C. The samples were allowed to rest for a week before being investigated by light scattering.* **(b)** *Ionic strength dependence of the hydrodynamic diameter measured on dispersions containing $PAA_{2K}-\gamma-Fe_2O_3$ particles and oppositely charged polymers (Z = 1). The ionic strength was monitored by addition of ammonium chloride ($NH_4Cl$). The color code is the same as in a). With decreasing $I_S$ the desalting transition occurred at a critical ionic strength 0.37 M, 0.54 M and 2.3 M for $PTEA_{11K}-b-PAM_{30K}$, PDADMAC and PEI respectively. The empty symbols and hatched area indicate hydrodynamic sizes that could not be measured by light scattering. Optical microscopy was used instead.*

The desalting transition was shown to be a key feature in the fabrication process of the nanowires; its occurrence with widespread homoPEs had then to be investigated. The ionic strength of mixed dispersions containing $PAA_{2K}-\gamma-Fe_2O_3$ and PDADMAC, PEI and PAH was increased up to 3 M ($NH_4Cl$) to verify the efficiency of the screening. Effective screening was found for PDADMAC and PEI, but not for PAH. For this later system, even at 3 M, the oppositely charged species interacted strongly and large aggregates were formed ($D_H = 400$ nm, SI-3). Fig. 2b displays the desalting transition for $PAA_{2K}-\gamma-$

Fe$_2$O$_3$/PDADMAC and PAA$_{2K}$–γ-Fe$_2$O$_3$/PEI, at the critical ionic strengths 0.54 M and 2.30 M respectively. With the homoPEs, the features of the desalting transition remained similar to those found with copolymers. The sizes of the clusters below the critical value increased however rapidly with decreasing I$_S$. As anticipated, Fig. 2b suggests that the interactions are stronger with homoPEs than with the diblock. Dispersions prepared apart from the charge stoichiometry *i.e.* at Z = 0.3 and Z = 7 were found to undergo similar desalting transitions, with aggregates of smaller sizes and slightly shifted ionic strengths (SI-4).

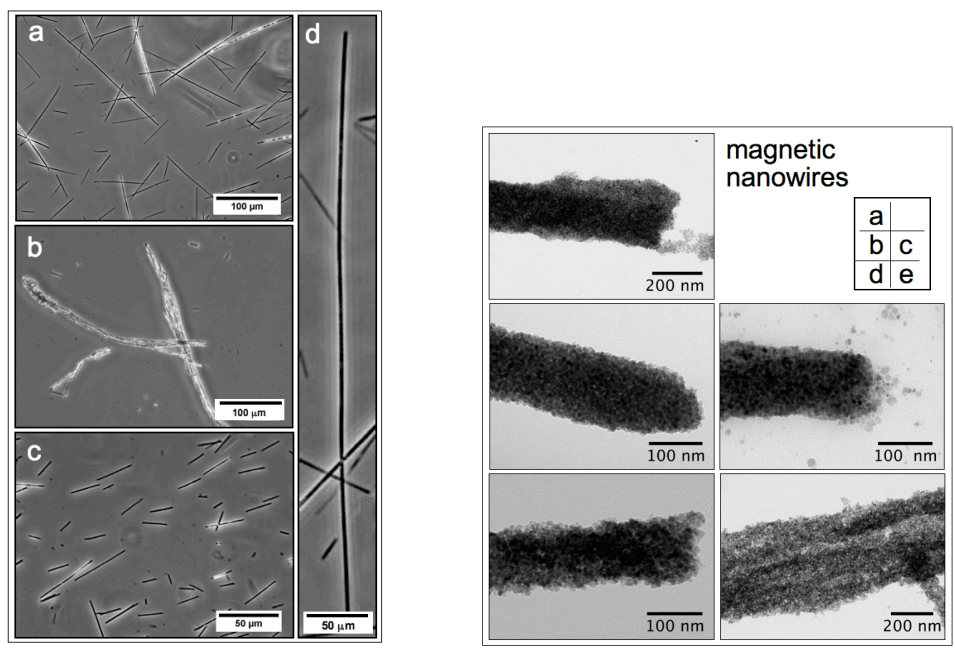

**Figure 3 :** *Phase-contrast optical microscopy images (a, b : ×20, c : ×40, d : ×10) of a dispersion of nanostructured wires obtained by dialysis from 8.3 nm γ-Fe$_2$O$_3$ particles and PDADMAC at Z = 0.3 (a,d), Z = 1 (b) and Z = 7 (c). The wires appear thicker in d) because of the use of a lower resolution objective. The images were taken in the absence of magnetic field.*

**Figure 4 :** *TEM images of nanostructured wires made from PAA$_{2K}$–γ-Fe$_2$O$_3$ nanoparticles and polymers : a) PTEA$_{11K}$–b–PAM$_{30K}$ (Z = 1), b) PDADMAC (Z = 0.3); c) PDADMAC (Z = 7); d) PEI (Z = 0.3) and e) PEI (Z = 7). The diameters of the wires vary between 150 and 400 nm.*

## III.3 − Desalting transition in the presence of a magnetic field

From the initial 3 cationic polymers, only PDADMAC and PEI exhibited a well-marked desalting transition. These polymers were tested with respect to dialysis in the presence of a 0.3 T magnetic field to stimulate the unidirectional growth of the aggregates (Fig. 1). Once the ionic strength of the dialysis bath reached its stationary value, the magnetic field was removed and the solutions were studied by optical microscopy. Figs. 3 show optical transmission microscopy images of aggregates made of PAA$_{2K}$–γ-Fe$_2$O$_3$/PDADMAC dispersions at $Z = 0.3$ (Fig. 3a), 1 (Fig. 3b) and 7 (Fig. 3c). Large and irregular aggregates in the 100 µm range were obtained at $Z = 1$, illustrating again that at the charge stoichiometry the interactions were strong and that electrostatic aggregates grew uncontrolled. With an excess of cationic ($Z = 0.3$) or anionic ($Z = 7$) charges however, regular nanostructured wires were readily fabricated. The wires grown apart from the charge stoichiometry ($Z = 0.3$ and 7) had a median length $L_0 = 90$ µm and 20 µm respectively, and exhibited the same mechanical and magnetic properties than the ones made from $PTEA_{11K}$-$b$-$PAM_{30K}$ copolymers : they were all rigid threads with superparamagnetic properties inherited from the single particles [22,36]. As an illustration of their magnetic properties, a movie of wires subjected to $\pi/2$-reorientations is shown in Supporting Information (Movie#1). The results obtained with PDADMAC were utterly reproduced with poly(ethyleneimine) (SI-5), showing *in fine* that the wire formation is a general phenomenon that does not depend on the nature of the polycations. For all polymers tested, the length distributions of the wires were found to be log-normal with median length $L_0$ (= 10 – 100 µm) and polydispersity 0.5. With PDADMAC, the tail of the distribution contained few but very long wires of length ½ mm, suggesting the possible influence of the chain architecture (linear for PDADMAC and branched for PEI) on the wire formation (Fig. 3d). To confirm that the wires made with copolymers and homoPEs had the same structures, TEM was monitored on PDADMAC and PEI-based dispersions. Figs. 4 compare the internal structures of wires obtained with $PTEA_{11K}$-$b$-$PAM_{30K}$ at $Z = 1$ (Fig. 4a), PDADMAC at $Z = 0.3$ and 7 (Figs. 4b and 4c) and PEI at $Z = 0.3$ and 7 (Figs. 4d and 4e) as binding agents. Extremities or intermediate pieces reveal diameters comprised between 150 and 400 nm with similar particles densities. These findings are in good agreement with the optical microscopy images which displayed no major differences between wires at the micron size. Note finally that the desalting transition technique allows the high-throughput synthesis, typically $10^{10}$ wires of length 10 µm within 10 minutes.

In a recent paper[36], it was concluded that the mechanism for the wire formation proceeded in two steps : *i)* the nucleation and growth of spherical clusters of particles, and *ii)* the alignment of the clusters induced by the magnetic dipolar interactions. For the kinetics, the clusters growth and their alignment occurred in parallel, leading to a continuous welding of the cylindrical structure. The results obtained with homoPEs indicate that the kinetics of co-assembly at charge stoichiometry is extremely rapid. All three pathways explored, *direct mixing* (Fig. 2a), *dilution* (Fig. 2b) and *dialysis* (Fig. 3) showed it : at $Z = 1$, the nanoparticle aggregates are the largest and their morphologies mainly uncontrolled. With block copolymers, the co-assembly kinetics was slowed down by the presence of the inert neutral blocks. With homoPEs, the finding of regular nanowires apart from stoichiometry suggests a kinetics governed by the balance between electrostatic repulsions and dipolar attractions between the precursor aggregates. We thus anticipate that the wires made with an excess of polymers should be positively charged and those with an excess of nanoparticles negatively charged. Electrokinetic measurements were performed on the wire dispersions to determine their electrostatic charges, however the data remained inconclusive because of the large sizes of the aggregates. Smaller Brownian isotropic clusters ($D_H = 250$ nm) were thus fabricated using the same dialysis conditions but without external magnetic field. Electrophoretic mobility and $\zeta$–potential were determined for PDADMAC and PEI based complexes at $Z = 0.3$, 1 and 7 using Zetasizer Nano ZS Malvern Instrument. The intensity distribution of the electrophoretic mobility shown in Fig. 5 for the three systems confirmed our hypothesis. At $Z = 0.3$, $\mu_E$ is centered around $+3 \times 10^{-4}$ cm$^2$ V$^{-1}$ s$^{-1}$ for both polyelectrolytes, whereas it is approximately 0 at $Z = 1$ and $-2.2 \times 10^{-4}$ cm$^2$ V$^{-1}$ s$^{-1}$ at $Z = 7$ (SI-6). A final test was performed to confirm these results. Positively ($L_0 = 90$ µm) and negatively ($L_0 = 20$ µm) charged PDADMAC wires were mixed together. The suspended colloids precipitated, resulting in the formation of large bottlebrush-like aggregates (Fig. 5c), where the short wires agglutinated onto the large ones thanks to attractive electrostatic interactions, suggesting oppositely charged wires.

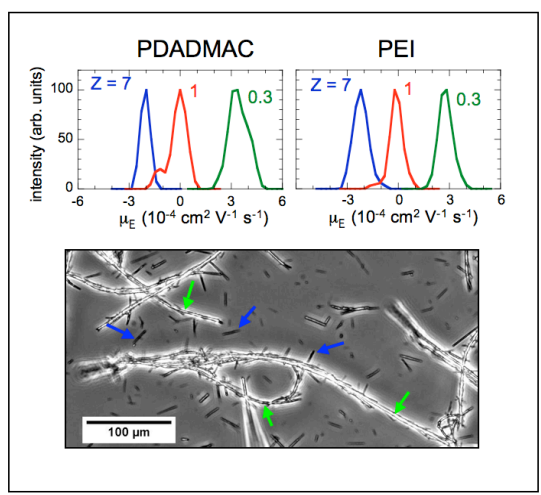

**Figure 5 -** *Upper panels : Intensity versus electrophoretic mobility for PAA$_{2K}$–γ-Fe$_2$O$_3$/PDADMAC and PAA$_{2K}$–γ-Fe$_2$O$_3$/PEI isotropic clusters obtained by dialysis. Lower panel : Phase-contrast optical microscopy images (×20) of a dispersion of positively (green arrows) and negatively (blue arrows) charged wires made from PAA$_{2K}$–γ-Fe$_2$O$_3$ and PDADMAC.*

## IV - Conclusion

In conclusion, we have shown that the previous copolymer based co-assembly strategy could be generalized to strong and weak polyelectrolytes. In terms of cost and practicality, this represents a remarkable improvement. Beyond, it shows that specific polymer architectures are not essential for the elaboration of supracolloidal electrostatic assemblies of adaptable size, charge and morphology, the structural, mechanical and magnetic properties being independent of the nature of the binding agents. The enhanced stability of the nanostructures built up with anionic iron oxide nanoparticles and cationic homoPEs is interpreted as resulting from a non-equilibrium associating process. Once the aggregates are formed, their microstructure remained unchanged over a period longer than years. As a result, the wires can be moved in the solvent or rotated by the application of a field or gradient without changing their internal structure. Because the present strategy for electrostatic co-assembly of nanoparticles is simple and versatile, it should open new perspectives for the design of nanodevices, such as tips, tweezers, actuators applicable in biophysics (e.g. bio-microrheology) and biomedicine for stimulating and sorting living cells or as novel contrast agents. We envision very exciting applications in these domains where the wires can be tailored on-demand according to the medium under scrutiny.

## Acknowledgement

This research was supported in part by Rhodia (France), by the Agence Nationale de la Recherche under the contracts BLAN07-3_206866 and ANR-09-NANO-P200-36, by the European Community through the project : "NANO3T—Biofunctionalized Metal and Magnetic Nanoparticles for Targeted Tumor Therapy", project number 214137 (FP7-NMP-2007-SMALL-1) and by the Région Ile-de-France in the DIM framework related to Health, Environnement and Toxicology (SEnT).

## Supporting Information

The Supporting Information section provides the characterization of the iron oxide nanoparticles and that of the cationic polymers. The evolution of the hydrodynamic diameter of particles and polymers as a function of ionic strength as well as the phenomonology of the desalting transition are described in details. Magnetic nanowires obtained from PEI homopolyelectrolytes and results of the electrophoretic mobility and ζ-potential are also shown. Movie#1 shows a series of reorientations of wires following 90°-flips of the magnetic field. This information is available free of charge via the Internet at http://pubs.acs.org/.

## References

(1)    Safarik, I.; Safarikova, M. *Journal of Chromatography B* **1999**, *722*, 33-53.

(2)    Hultgren, A.; Tanase, M.; Felton, E. J.; Bhadriraju, K.; Salem, A. K.; Chen, C. S.; Reich, D. H. *Biotechnology Progress* **2005**, *21*, 509 - 515.

(3)     Fung, A. O.; Kapadia, V.; Pierstorff, E.; Ho, D.; Chen, Y. *The Journal of Physical Chemistry C* **2008**, *112*, 15085-15088.

(4)     Johansson, F.; Jonsson, M.; Alm, K.; Kanje, M. *Experimental Cell Research* **2010**, *316*, 688-694.

(5)     Song, M. M.; Song, W. J.; Bi, H.; Wang, J.; Wu, W. L.; Sun, J.; Yu, M. *Biomaterials* **2010**, *31*, 1509-1517.

(6)     Cappallo, N.; Lapointe, C.; Reich, D. H.; Leheny, R. L. *Physical Review E* **2007**, *76*, 6.

(7)     Goubault, C.; Jop, P.; Fermigier, M.; Baudry, J.; Bertrand, E.; Bibette, J. *Physical Review Letters* **2003**, *91*, 260802.

(8)     Biswal, S. L.; Gast, A. P. *Physical Review E* **2003**, *68*, 021402.

(9)     Cebers, A. *Current Opinion in Colloid & Interface Science* **2005**, *10*, 167-175.

(10)    Lee, D.; Cohen, R. E.; Rubner, M. F. *Langmuir* **2006**, *23*, 123-129.

(11)    Fert, A.; Piraux, L. *Journal of Magnetism and Magnetic Materials* **1999**, *200*, 338-358.

(12)    Hurst, S. J.; Payne, E. K.; Qin, L. D.; Mirkin, C. A. *Angew. Chem.-Int. Edit.* **2006**, *45*, 2672-2692.

(13)    Furst, E. M.; Suzuki, C.; Fermigier, M.; Gast, A. P. *Langmuir* **1998**, *14*, 7334.

(14)    Singh, H.; Laibinis, P. E.; Hatton, T. A. *Nano Letters* **2005**, *5*, 2149-2154.

(15)    Sun, S.; Anders, S.; Hamann, H. F.; Thiele, J.-U.; Baglin, J. E. E.; Thomson, T.; Fullerton, E. E.; Murray, C. B.; Terris, B. D. *Journal of the American Chemical Society* **2002**, *124*, 2884-2885.

(16)    Ge, J.; Hu, Y.; Yin, Y. *Angewandte Chemie International Edition* **2007**, *46*, 7428 - 7431.

(17)    Srivastava, S.; Samanta, B.; Jordan, B. J.; Hong, R.; Xiao, Q.; Tuominen, M. T.; Rotello, V. M. *Journal of the American Chemical Society* **2007**, *129*, 11776-11780.

(18)    Stolarczyk, J. K.; Ghosh, S.; Brougham, D. F. *Angew. Chem.-Int. Edit.* **2009**, *48*, 175-178.

(19)    Sheparovych, R.; Sahoo, Y.; Motornov, M.; Wang, S. M.; Luo, H.; Prasad, P. N.; Sokolov, I.; Minko, S. *Chemistry of Materials* **2006**, *18*, 591 - 593.

(20)    Park, J.-H.; Maltzahn, G. v.; Zhang, L.; Michael P. Schwartz; Ruoslahti, E.; Bhatia, S. N.; Sailor, M. J. *Advanced Materials* **2008**, *20*, 1630–1635.

(21)    Yuwono, V. M.; Burrows, N. D.; Soltis, J. A.; Penn, R. L. *Journal of the American Chemical Society* **2010**, *132*, 2163-2165.

(22)    Fresnais, J.; Berret, J.-F.; Frka-Petesic, B.; Sandre, O.; Perzynski, R. *Adv. Mater.* **2008**, *20*, 3877-3881.

(23)    Fresnais, J.; Lavelle, C.; Berret, J.-F. *The Journal of Physical Chemistry C* **2009**, *113*, 16371-16379.

(24)    Massart, R.; Dubois, E.; Cabuil, V.; Hasmonay, E. *J. Magn. Magn. Mat.* **1995**, *149*, 1 - 5.

(25)    Bee, A.; Massart, R.; Neveu, S. *J. Magn. Magn. Mat.* **1995**, *149*, 6 - 9.

(26)    Berret, J.-F.; Sandre, O.; Mauger, A. *Langmuir* **2007**, *23*, 2993-2999.

(27)    Jacquin, M.; Muller, P.; Talingting-Pabalan, R.; Cottet, H.; Berret, J.-F.; Futterer, T.; Theodoly, O. *J. Colloid Interface Sci.* **2007**, *316*, 897-911.

(28)    Berret, J.-F. *Macromolecules* **2007**, *40*, 4260-4266.

(29)    Bertrand, P.; Jonas, A.; Laschewsky, A.; Legras, R. *Macromol. Rapid Commun.* **2000**, *21*, 319-348.

(30)    Ostrander, J. W.; Mamedov, A. A.; Kotov, N. A. *J. Am. Chem. Soc.* **2001**, *123*, 1101 - 1110.

(31)     Tranchant, I.; Thompson, B.; Nicolazzi, C.; Mignet, N.; Scherman, D. *J. Gene. Med.* **2004**, *6*, S24-S35.

(32)     Bordi, F.; Cametti, C.; Diociaiuti, M.; Gaudino, D.; Gili, T.; Sennato, S. *Langmuir* **2004**, *20*, 5214-5222.

(33)     Burgh, S. v. d.; Keizer, A. d.; Stuart, M. A. C. *Langmuir* **2004**, *20*, 1073 - 1084.

(34)     Qi, L.; Fresnais, J.; Berret, J.-F.; Castaing, J. C.; Grillo, I.; Chapel, J. P. *Journal of Physical Chemistry C* **2010**, *114*, 12870-12877.

(35)     Qi, L.; Fresnais, J.; Berret, J.-F.; Castaing, J.-C.; Destremaut, F.; Salmon, J.-B.; Cousin, F.; Chapel, J.-P. *The Journal of Physical Chemistry C* **2010**, *114*, 16373-16381.

(36)     Yan, M.; Fresnais, J.; Berret, J.-F. *Soft Matter* **2010**, *6*, 1997-2005.